\documentclass[twocolumn]{jpsj2} 
\title{Magnetic-Susceptibility and Specific-Heat Studies on the Inhomogeneity of Superconductivity in the Underdoped La$_{2-x}$Sr$_x$CuO$_4$}

\author{\textsc{Tadashi Adachi}\thanks{E-mail address: adachi@teion.apph.tohoku.ac.jp}, \textsc{Keisuke Omori}, \textsc{Yoichi Tanabe} and \textsc{Yoji Koike}}

\inst{Department of Applied Physics, Graduate School of Engineering, Tohoku University, \\6-6-05 Aoba, Aramaki, Aoba-ku, Sendai 980-8579}

\abst{The possible inhomogeneity of superconductivity has been investigated by means of {\it bulk-sensitive probes}, using La$_{2-x}$Sr$_x$CuO$_4$ (LSCO) single crystals from the underdoped to optimally doped regime. 
Measurements of the magnetic susceptibility, $\chi$, on field cooling and specific heat have revealed that both the absolute value of $\chi$ at 2 K, regarded as corresponding to the superconducting (SC) volume fraction in a sample, and the Sommerfeld constant, reflecting the density of states of quasiparticles at the Fermi level, exhibit significant dependence on $x$, namely, on the hole concentration. 
This is the first experimental work strongly suggesting that a phase separation into SC and normal-state regions takes place in the underodped regime of LSCO as well as in the overdoped regime.}

\kword{phase separation, La$_{2-x}$Sr$_x$CuO$_4$, magnetic susceptibility, specific heat, superconducting volume fraction, Sommerfeld constant}

\begin{document}
\maketitle

\section{Introduction}\label{intro}
It has been elucidated in recent years that electronic inhomogeneity is one of salient features in the high-$T_{\rm c}$ cuprates~\cite{pan}. 
The electronic inhomogeneity may be an intrinsic nature in the CuO$_2$ plane related to an electronic ordered state such as one-dimensional stripes of holes and spins~\cite{tranquada} or two-dimensional square-lattice charge ordering,~\cite{hanaguri,ido} rather than caused by the randomness of dopants in a crystal. 
In the overdoped regime of the hole-doped high-$T_{\rm c}$ cuprates, in fact, spatial inhomogeneity, namely, a phase separation into superconducting (SC) and normal-state regions has been suggested from measurements of the muon spin relaxation ($\mu$SR) in Tl$_2$Ba$_2$CuO$_{6+\delta}$ (TBCO), (Y,Ca)Ba$_2$Cu$_3$O$_{7-\delta}$ (YCBCO) and Tl$_{1-y}$Pb$_y$Sr$_2$Ca$_{1-x}$Y$_x$Cu$_2$O$_7$~\cite{uemura,niedermayer,bernhard}, specific heat in La$_{2-x}$Sr$_x$CuO$_4$ (LSCO) and TBCO~\cite{loram,loram2,wen}, magnetic susceptibility, $\chi$, in LSCO, TBCO, YCBCO, Bi$_2$Sr$_2$CaCu$_2$O$_{8+\delta}$ (BSCCO), Bi$_2$Sr$_2$CuO$_{6+\delta}$ (BSCO)~\cite{wen2,tanabe,adachi,tanabe2}, nuclear magnetic resonance in LSCO~\cite{ohsugi} and scanning tunneling microscopy and scanning tunneling spectroscopy (STM/STS) in BSCO~\cite{boyer}. 

In the underdoped high-$T_{\rm c}$ cuprates, on the other hand, STM/STS measurements in BSCCO have revealed spatial inhomogeneity of the energy gap at the Fermi level at low temperatures, suggesting a spatially inhomogeneous distribution of the SC gap and the so-called pseudo gap~\cite{lang}. 
Recently, one-dimensional stripe-like charge modulations have been observed in the pseudo-gapped region of the spatially inhomogeneous state in Ca$_{2-x}$Na$_x$CuO$_2$Cl$_2$ and Bi$_2$Sr$_2$Ca$_{1-x}$Dy$_x$Cu$_2$O$_{8+\delta}$~\cite{kohsaka,kohsaka2}. 
Specific-heat measurements have revealed that the so-called Sommerfeld constant in the ground state, $\gamma$, is not zero even in SC samples of LSCO~\cite{kato,momono,nohara,wen3}, suggesting the presence of normal-state regions in the SC samples and therefore supporting the spatially inhomogeneous picture of superconductivity. 
Moreover, $\mu$SR~\cite{niedermayer2,sanna,ishida} and NMR/NQR~\cite{julien} in the underdoped LSCO, YCBCO and YBa$_2$Cu$_3$O$_{7-\delta}$ (YBCO) measurements have revealed the existence of a spin-glass-like static magnetic state, suggesting the coexistence of SC and static magnetic states. 
From the theoretical viewpoint, the inhomogeneous state in the underdoped high-$T_{\rm c}$ cuprates has been argued as being due to a spatially modulated state of $d$-wave superconductivity and spin stripe order~\cite{ogata}, or due to a phase separation into antiferromagnetically ordered, SC and non-ordered regions~\cite{dagotto}, or due to a granular SC state~\cite{yanase}. 

In this paper, we aim at investigating the possible inhomogeneity of superconductivity using {\it bulk-sensitive probes} reflecting bulk properties of a sample, which are not influenced by any possible peculiar surface state. 
We have grown high-quality single crystals of LSCO from the underdoped to optimally doped regime and estimated the SC volume fraction from the $\chi$ measurements on field cooling~\cite{omori,koike}, which was the same way carried out in the overdoped LSCO~\cite{tanabe,adachi}. 
Furthermore, using the same single crystals of good quality used for the $\chi$ measurements, the specific heat has been measured to clarify the systematic hole-concentration dependence of $\gamma$ in the ground state~\cite{koike}. 
The use of high-quality single crystals is essential for the study of the possible inhomogeneous electronic state, because the inhomogeneity due to crystal imperfections must be reduced as much as possible.

\section{Experimental details}
\begin{figure}[tbp]
\begin{center}
\includegraphics[width=1.0\linewidth]{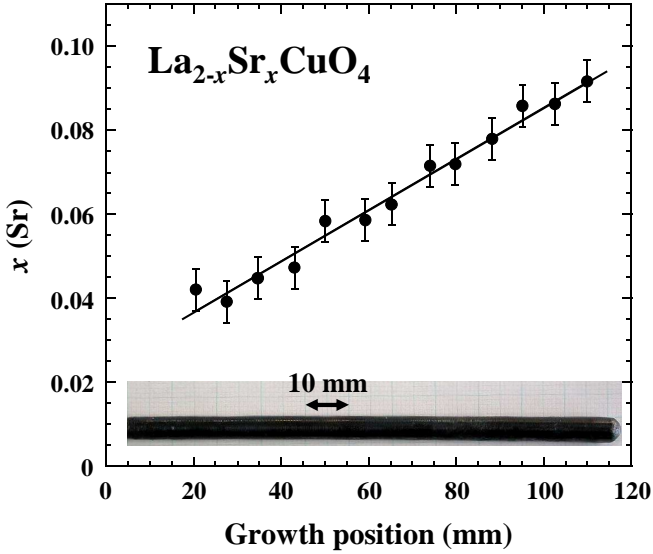}
\end{center}
\caption{(Color online) Sr-concentration, $x$, estimated from ICP-OES vs. the growth position of a single-crystal rod of La$_{2-x}$Sr$_x$CuO$_4$ with $x=0.04-0.10$. Note that the scale of the horizontal axis corresponds to that of the photo of an as-grown single-crystal rod of Sr-concentration-gradient La$_{2-x}$Sr$_x$CuO$_4$ in the inset.}  
\label{icp} 
\end{figure}

Sr-concentration-gradient single crystals of LSCO were grown by the traveling-solvent floating-zone method under flowing O$_{\rm 2}$ gas of 4 bar. 
Advantages of the use of the Sr-concentration-gradient single crystals are that plenty of samples with various $x$ values are able to be taken from a single-crystal rod, keeping the crystallinity identical with each other, and that fine Sr-concentration dependence of physical properties is able to be investigated. 
The details of the preparation of powders for the feed rod and solvent are almost similar to those formerly reported~\cite{kawamata,tanabe} 
For the feed rod, stoichiometric powders of LSCO with $x=0.04-0.18$ at an interval of 0.01 were prepared. 
The value of $x$ was changed every $1.5-2.0$ cm at an interval of 0.01 in a feed rod.
Using feed rods with $x=0.04-0.10$, $0.09-0.15$, $0.11-0.16$, $0.14-0.18$, four single-crystal rods were grown at a rate of 0.7 mm/h, so that the gradient of the Sr concentration in each rod became gentle. 
The as-grown single-crystal rods were annealed in flowing O$_{\rm 2}$ gas of 1 bar at 900$^{\rm o}$C for 50 h, cooled down to 500$^{\rm o}$C at a rate of 8$^{\rm o}$C/h, kept at 500$^{\rm o}$C for 50 h and then cooled down to room temperature at a rate of 8$^{\rm o}$C/h.

Samples, obtained by slicing the single-crystal rods almost perpendicularly to the direction of the crystal growth to form many pieces of 1 mm in thickness, were formed into the same rectangular shape of $2.9 \times 1.0$ mm$^2$ in the ab plane and 1.2 mm along the c-axis within the error of $\pm$4 \%, in order to make the demagnetizing-field effect identical among the samples in the $\chi$ measurements.
The Sr content of each sample was analyzed by inductively-coupled-plasma optical-emission-spectrometry (ICP-OES). 
The quality of the samples was checked by x-ray back-Laue photography to be good. 
The samples were also checked by powder x-ray diffraction, in which no Bragg peaks due to impurities were observed. 
The full-width at half maximum of the (006) rocking curve was $0.1-0.2$ $^{\rm o}$, which is comparable to that reported in the previous literature~\cite{tanabe,komiya} and no systematic $x$ dependence was observed. 
The distribution of the Sr content in a sample was also checked using an electron probe microanalyzer (EPMA) to be homogeneous within the experimental accuracy.

The $\chi$ measurements were carried out in a magnetic field of 10 Oe parallel to the c-axis on both zero-field cooling and field cooling at low temperatures down to 2 K, using a superconducting quantum interference device magnetometer (Quantum Design, MPMS). 
The specific heat was measured by the thermal-relaxation method at low temperatures down to 2 K, using a commercial apparatus (Quantum Design, PPMS).

\section{Results}\label{results}
The inset of Fig. 1 displays an as-grown single-crystal rod of Sr-concentration-gradient LSCO grown using a feed rod with $x=0.04-0.10$. 
The size is 5 mm in diameter and 110 mm in length. 
As shown in Fig. 1, the value of $x$ increases almost linearly in the direction of the crystal growth. 
Therefore, it is concluded that the growth of a Sr-concentration-gradient single crystal has been successful. 
The distribution of $x$, $\Delta x$, in one sample obtained by slicing the single-crystal rod at an interval of 1 mm was expected to be less than 0.0007, which is small enough to investigate the $x$ dependence of the following physical properties.

\begin{figure}[tbp]
\begin{center}
\includegraphics[width=0.9\linewidth]{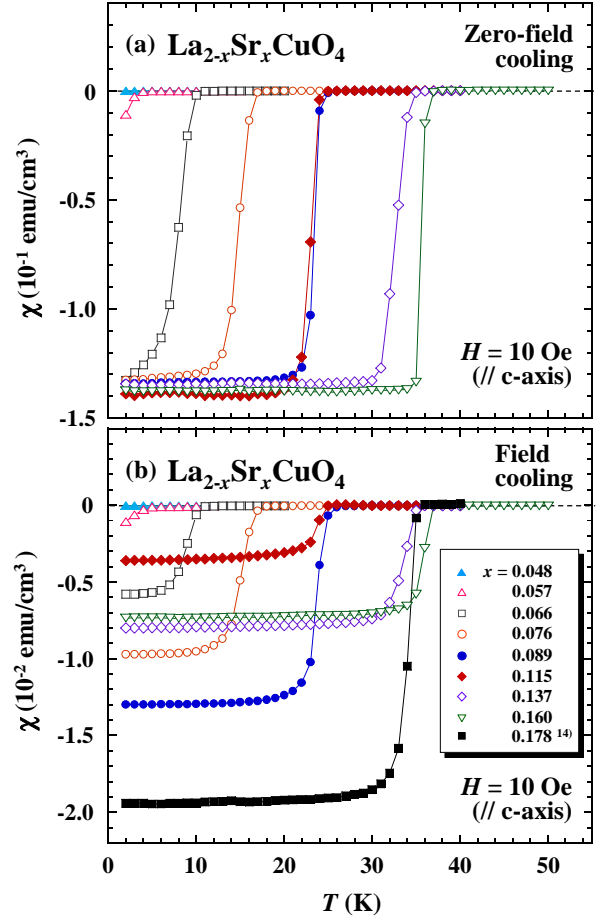}
\end{center}
\caption{(Color online) Temperature dependence of the magnetic susceptibility, $\chi$, for typical values of $x$ in La$_{2-x}$Sr$_x$CuO$_4$ with $x=0.048-0.160$ in a magnetic field of 10 Oe parallel to the c-axis (a) on zero-field cooling and (b) on field cooling. The data of $x=0.178$ on field cooling are shown for comparison.~\cite{tanabe2}}  
\label{chi} 
\end{figure}

\begin{figure}[tbp]
\begin{center}
\includegraphics[width=0.85\linewidth]{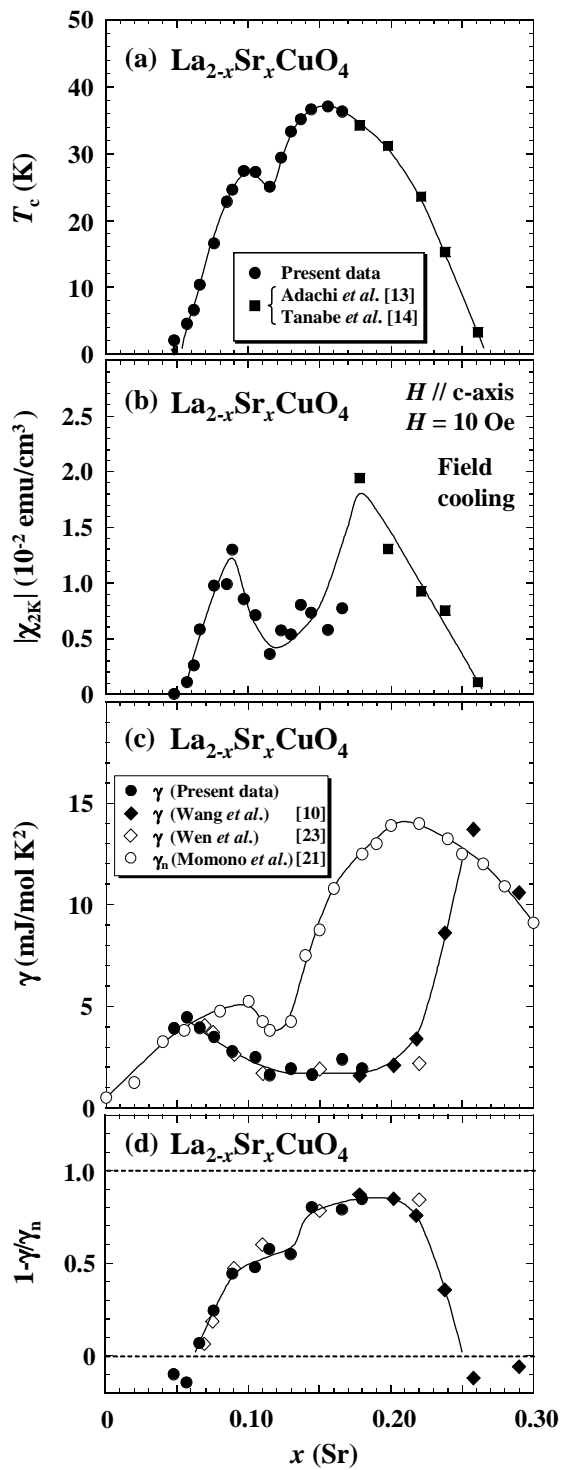}
\end{center}
\caption{Sr-concentration, $x$, dependences of (a) the superconducting transition temperature, $T_{\rm c}$, defined as the cross point between the extrapolated line of the steepest part of the Meissner diamagnetism and zero susceptibility, (b) the absolute value of $\chi$ at 2 K on field cooling, $|\chi_{\rm 2K}|$, in a magnetic field 10 Oe parallel to the c-axis, (c) the Sommerfeld constant, $\gamma$, (d) the ratio of $\gamma$ to the normal-state value of $\gamma$, $\gamma_{\rm n}$, obtained by Momono {\it et al}.~\cite{momono} using polycrystalline samples, $1-\gamma / \gamma_{\rm n}$, for La$_{2-x}$Sr$_x$CuO$_4$ single crystals. The data for $x \ge 0.178$ in (a) and (b) have been obtained using our single crystals and are also plotted for comparison.~\cite{adachi,tanabe2} Values of $\gamma$ and $1-\gamma / \gamma_{\rm n}$ obtained by Wang {\it et al}.~\cite{wen} using our single crystals and by Wen {\it et al}.~\cite{wen3} using single crystals are also plotted for comparison in (c) and (d). Values of $\gamma_{\rm n}$ obtained by Momono {\it et al}.~\cite{momono} using polycrystalline samples are also plotted in (c). Solid lines are to guide the reader's eye.}  
\label{SCvf} 
\end{figure}

Figure 2(a) shows the temperature dependence of $\chi$ on zero-field cooling for typical values of $x$ in LSCO. 
A sharp SC transition is observed in each $x$, indicating the good quality of the samples. 
The SC transition temperature, $T_{\rm c}$, increases with increasing $x$ up to $x=0.160$ but is slightly suppressed for $x=0.115$ due to the so-called 1/8 anomaly. 
It is found that values of $\chi$ at 2 K are almost identical among samples of $x \ge 0.066$, suggesting that the effect of the demagnetizing field on $\chi$ is identical among the samples as we expected. 
For $x<0.066$, on the other hand, the absolute value of $\chi$ at 2 K is found to be small, because the applied field of 10 Oe is larger than the lower critical field at 2 K. 
This has been confirmed from the magnetization measurements in low magnetic fields. 

Figure 2(b) shows the temperature dependence of $\chi$ on field cooling for typical values of $x$ in LSCO. 
It is found that the absolute value of $\chi$ at 2 K on field cooling, $|\chi_{\rm 2K}|$, increases with increasing $x$ up to $x=0.089$, decreases and exhibits the minimum at $x=0.115$ and increases again up to $x=0.178$.~\cite{tanabe2} 

The Sr-concentration dependence of $T_{\rm c}$, defined as the cross point between the extrapolated line of the steepest part of the Meissner diamagnetism and zero susceptibility, is shown in Fig. 3(a). 
The $T_{\rm c}$ exhibits a parabolic change and is slightly suppressed around $x=0.115$, which is consistent with the previous report.~\cite{koike-ssc}
The Sr-concentration dependence of $|\chi_{\rm 2K}|$ is shown in Fig. 3(b), together with our former results in the overdoped regime of $x \ge 0.178$ indicating a monotonic decrease with increasing $x$.~\cite{tanabe,adachi,tanabe2} 
The significant dependence of $|\chi_{\rm 2K}|$ on $x$ suggests the change of the vortex-pinning effect in a sample depending on $x$. 
The decrease in $|\chi_{\rm 2K}|$ with increasing $x$ in the overdoped regime of $x>0.178$ has already been concluded to be due to a phase separation into SC and normal-state regions in our previous papers.~\cite{tanabe,adachi,tanabe2} 

\begin{figure}[tbp]
\begin{center}
\includegraphics[width=1.0\linewidth]{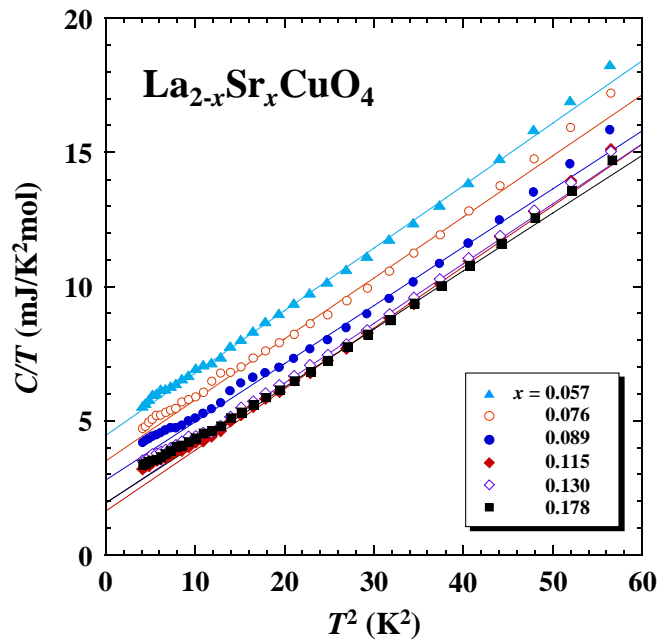}
\end{center}
\caption{(Color online) Temperature dependence of the specific heat, $C$, for typical values of $x$ in La$_{2-x}$Sr$_x$CuO$_4$ with $x=0.057-0.178$ plotted as $C/T$ vs. $T^2$. Solid lines indicate $C/T = \gamma + \beta T^2$ using $\gamma$ and $\beta$ values obtained from the best-fit results using Eq. (1).} 
\label{SH} 
\end{figure}

Figure 4 shows the temperature dependence of the specific heat, $C$, for typical values of $x$ in LSCO plotted as $C/T$ vs. $T^2$. 
To extract the electronic contribution from the total specific heat, the data were fitted to the following equation.
\begin{equation}
C(T) = \gamma T + A T^2 + \beta T^3 + \frac{D}{T^2}.
\label{eq1}
\end{equation}
The first term represents the electronic specific heat residing in the heart of the present study. 
The value of $\gamma$ is proportional to the density of states of quasiparticles at the Fermi level and therefore is expected to be zero in the ground state of a homogeneous SC sample. 
In an inhomogeneous state, on the other hand, $\gamma$ is expected to be finite even in the ground state due to the creation of normal-state regions in a sample. 
The second term is known as the $d$-wave SC one due to the excitation of quasiparticles around the node of the SC gap around ($\pi/2$, $\pi/2$) in $k$ space. 
The value of $A$ has been reported to be negligibly small in the underdoped regime.~\cite{momono,wen3}
In fact, no physically meaningful finite values of $A$ were obtained from the present study and therefore the second term was omitted in the present fitting. 
The third term represents the phonon specific heat. 
The final is the Schottky term, although the origin is not clear. 
This may be due to a very small amount of paramagnetic impurities.~\cite{nohara}

Solid lines in Fig. 4 indicate $C/T = \gamma + \beta T^2$ using $\gamma$ and $\beta$ values obtained from the best fit of the data with Eq. (1). 
It is found that the $\gamma$ value is finite in the ground state even in SC samples. 
Moreover, $\gamma$ appears to decrease with increasing $x$ and tends to be saturated around the optimally doped regime. 
The Sr-concentration dependence of $\gamma$ is shown in Fig. 3(c), together with the normal-state values of $\gamma$, $\gamma_{\rm n}$, obtained by Momono {\it et al}.~\cite{momono} using polycrystalline samples. 
The values of $\gamma$ are almost consistent with those formerly reported.~\cite{wen3} 
It is found for the underdoped and optimally doped regimes of $x=0.057-0.178$ that $\gamma$ exhibits a finite value even in the SC state as well as in the overdoped regime.~\cite{wen} 
With increasing $x$, $\gamma$ monotonically decreases and tends to be saturated around the optimally doped regime of $x=0.15-0.18$, followed by a rapid increase in the overdoped regime.~\cite{wen}
These results suggest that there exists a finite density of states at the Fermi level even in the ground state of superconductivity in LSCO.

\section{Discussion}
First, we discuss the significant $x$ dependence of $|\chi_{\rm 2K}|$ for LSCO shown in Fig. 3(b). 
As a factor to reduce the value of $|\chi_{\rm 2K}|$, the vortex-pinning effect due to crystal imperfections must be taken into account more or less. 
This is, however, negative as follows. 
Since Sr-concentration-gradient single crystals, in which the crystallinity was believed to be identical among the samples as confirmed from the (006) rocking curve, were used for the measurements, the significant $x$ dependence of $|\chi_{\rm 2K}|$ is not explained simply. 
Moreover, since the vortex-pinning effect is expected to be most pronounced around $x=0.18$ where the SC condensation energy exhibits the maximum,~\cite{matsuzaki} $|\chi_{\rm 2K}|$ should show the minimum around $x=0.18$ and increases monotonically with both increasing and decreasing $x$, but this is completely inconsistent with the present results. 

The vortex-pinning effect of twin boundaries must be taken into account also, because LSCO undergoes a structural phase transition from the tetragonal high-temperature phase (space group: $I4/mmm$) to orthorhombic mid-temperature phase ($Bmab$). 
The transition temperature decreases with increasing $x$ and disappears around $x=0.22$.~\cite{thurston,takagi} 
This indicates that the orthorhombicity at low temperatures decreases with increasing $x$, resulting in a gradual weakening of the vortex-pinning effect of twin boundaries and an enhancement of $|\chi_{\rm 2K}|$. 
This hypothesis is, however, incompatible with the decrease in $|\chi_{\rm 2K}|$ for $x>0.089$. 
In general, the vortex-pinning effect of twin boundaries often observed in YBCO is believed to be relatively weak compared with that of crystal imperfections in a bulk material.~\cite{twin} 
Therefore, the present $x$ dependence of $|\chi_{\rm 2K}|$ is unable to be reproduced in terms of the vortex-pinning effect of only twin boundaries, suggesting that it reflects the $x$ dependence of the SC volume fraction in a sample. 
In other words, the SC volume fraction probably increases with increasing $x$ for $x \le 0.089$. 

The decrease in $|\chi_{\rm 2K}|$ around $x=0.115$ might be related to the formation of the spin-charge stripe order.~\cite{tranquada}
Neutron-scattering~\cite{suzuki,kimura} and muon-spin-relaxation~\cite{nabe-prb,ada-prb} measurements have suggested that the static stripe order is formed at low temperatures around $x=0.115$, being concomitant with a slight decrease in $T_{\rm c}$. 
As the SC volume fraction is suppressed around $x=0.115$, it is reasonably understood that a phase separation into SC and stripe-ordered non-SC regions occurs in a sample around $x=0.115$. 
For the optimally doped regime, on the other hand, it is found in Fig. 3(b) that the SC volume fraction is still suppressed. 
Possible reasons are that the effect of the stripe order spreads out to the optimally doped regime and/or that the vortex-pinning effect of crystal imperfections and/or twin boundaries is marked due to the large SC condensation energy.~\cite{matsuzaki} 

Next, the $x$ dependence of $\gamma$ shown in Fig. 3(c) is discussed. 
There exists a possibility that the observed finite values of $\gamma$ in the SC state originate from the scattering of electron pairs by impurities, generating quasiparticles around the node of the $d$-wave SC gap. 
According to the theory on the so-called Volovik effect, the Doppler shift of the energy of quasiparticles around the node of the SC gap by the application of magnetic field induces a finite value of $\gamma$ proportional to the square root of the magnetic field, $H$, in the clean limit, while the $\gamma$ value is proportional to $HlnH$ in the dirty limit.~\cite{volovik} 
It has been reported from the specific-heat measurements of LSCO in magnetic fields that $\gamma$ is enhanced with increasing $H$ in proportion to $H^{1/2}$ rather than $HlnH$,~\cite{wen4} indicating that the scattering of electron pairs by impurities is negligible. 
Furthermore, the number of quasiparticles around the node induced by the scattering between electron pairs and impurities is inversely proportional to the slope of the SC gap around the node. 
Recent photoemission-spectroscopy measurements of LSCO~\cite{hashimoto} and BSCCO~\cite{tanaka} have revealed that the size of the SC gap around the node is nearly unchanged between the optimally doped and underdoped regime of $p \sim 0.08$, suggesting that the slope of the SC gap around the node is nearly unchanged. 
Therefore, both the observed finite values of $\gamma$ in the SC state of the underdoped LSCO and the $x$ dependence of $\gamma$ are hardly understood as being due to the simple scattering of electron pairs by impurities, but they suggest the existence of normal quasiparticles in the ground state. 
In other words, both SC and normal carriers simultaneously exist in the ground state, being compatible with the phase-separation model deduced from the $\chi$ measurements. 
Accordingly, these results strongly suggest that a phase separation into SC and normal-state regions takes place in the underodped regime of LSCO, insisting on the occurrence of the phase separation as a bulk property. 
It is noted that $\gamma$ is small but finite around the optimally doped regime. 
This might be due to the overestimation of $\gamma$ caused by the neglect of the $d$-wave SC term in the specific heat, because it has been reported that the value of $A$ in Eq. (1) is small but increases with increasing $x$ in the underdoped regime.~\cite{momono} 

Here, we argue the relevance to the theories taking into account the inhomogeneous electronic state mentioned in Sec. 1. 
According to the model of a spatially modulated state of $d$-wave superconductivity and spin stripe order in the CuO$_2$ plane proposed by Himeda {\it et al}.,~\cite{ogata} sizes of the SC and non-SC regions are small compared with the size of vortices. 
Therefore, the non-SC regions are not expected to behave as effective pinning centers of vortices, so that the SC volume fraction is hardly suppressed. 
Accordingly, their model appears to be inconsistent with the present results. 
Yanase~\cite{yanase} has proposed a granular SC state in the underdoped regime in which the SC order parameter is suppressed around disordered areas in the CuO$_2$ plane due to large SC fluctuations. 
In this model, holes are homogenously distributed in the CuO$_2$ plane and $T_{\rm c}$ decreases with decreasing $p$ owing to the gradual degradation of the SC coherence between SC grains. 
It appears that this model well explains the present results and non-SC regions are regarded as normal-state regions formed around disordered areas. 
As for the phase-separation model by Alvarez {\it et al}.~\cite{dagotto} in which the effect of disorder is taken into account, a glassy state consisting of antiferromagnetically ordered, SC and non-ordered regions where values of $p$ are different from one another is formed in the CuO$_2$ plane in the underdoped regime. 
It appears that this is a kind of phase separation into SC and non-SC regions and consistent with the present results. 
However, the phase separation in both models by Yanase~\cite{yanase} and by Alvarez {\it et al}.~\cite{dagotto} is driven by disorder. 
If this is the case, $\gamma$ should increase in magnetic field in proportion to $HlnH$, which is inconsistent with the experimental result.~\cite{wen4} 
To test these theories, measurements in an underdoped high-$T_{\rm c}$ cuprate cleaner than LSCO are necessary.

The origin of the phase separation in the underdoped high-$T_{\rm c}$ cuprates has been under debate. 
The general understanding in the underdoped regime is due to the occurrence of strong-coupling superconductivity with a short SC coherence length.
The STM/STS measurements have actually revealed nano-scaled electronic inhomogeneity.~\cite{pan,lang} 
Therefore, it is a quite naive way of thinking that the phase separation in the underdoped regime originates from a large SC fluctuation. 
Moreover, it is possible that the scale of the phase separation is as microscopic as the SC coherence length of a few nanometers. 
As for the increase in $T_{\rm c}$ with increasing $p$ in the underdoped phase-separated state, it may be due to the gradual evolution of the SC coherence between SC regions with increasing $p$ because of the proximity effect or Josephson coupling. 
The situation may be similar to that observed in a granular thin film of lead overlaid by silver, whose $T_{\rm c}$ increases with increasing amount of silver due to the enhancement of the Josephson coupling between grains of lead.~\cite{lead} 
Note that, in the non-SC regions of the underdoped regime, a Fermi-liquid state is unlikely to be realized due to a poor number of holes, while it is likely in the non-SC regions of the overdoped regime.~\cite{tanabe} 
Since $\gamma$ is finite in the underdoped regime, it is speculated that the non-SC regions in the underdoped regime are not characterized by insulators but the Anderson localization due to a poor number of holes. 
From the magnetic point of view, a spin-glass-like state may be realized in the non-SC regions based on the $\mu$SR results.~\cite{niedermayer2,sanna,ishida}

Finally, we discuss the relation between observed values of $\gamma$ and the Fermi surface. 
Recent angle-resolved photoemission spectroscopy (ARPES) measurements have revealed that a gap opens in the antinodal region of the Fermi level around ($\pi$, 0) and (0, $\pi$) in $k$ space at high temperatures above $T_{\rm c}$, while the Fermi surface called a Fermi arc survives around the nodal region at high temperatures above $T_{\rm c}$ and opens a $d$-wave SC gap below $T_{\rm c}$.~\cite{damascelli} 
Where quasiparticles corresponding to the observed finite values of $\gamma$ reside on the Fermi surface is an intriguing issue which has not yet been clarified. 
Figure 3(d) shows the $x$ dependence of $1-\gamma / \gamma_{\rm n}$, that is, the ratio of the density of states at the Fermi level used for superconductivity to the total density of states at the Fermi level for LSCO. 
The value of $1-\gamma / \gamma_{\rm n}$ increases monotonically with increasing $x$ and becomes nearly unity around the optimally doped regime of $x \sim 0.18$, suggesting that most of carriers are condensed into the SC state in the ground state around the optimally doped regime. 
These indicate that, regardless of the size of the Fermi surface, the ratio of the SC carrier number to the total carrier number increases with doping of holes in the underdoped regime. 
Supposing that $\gamma_{\rm n}$ corresponds to the density of states at the Fermi arc suggested from recent ARPES results,~\cite{yoshida} the residual $\gamma$ value in the SC state ought to originate from the Fermi arc, which appears to be inconsistent with the ARPES results that a clear $d$-wave SC gap opens around the Fermi arc below $T_{\rm c}$. 
Accordingly, the concept of the phase separation must be taken into account and how effects of the phase separation in real space appears in $k$ space must be investigated.

\section{Summary}
We have investigated the possible phase separation into SC and normal-state regions, formerly suggested in the overdoped cuprates, by means of {\it bulk-sensitive} $\chi$ and specific-heat measurements for LSCO single crystals with $x=0.04-0.18$ covering the underdoped and optimally doped regimes. 
The $\chi$ measurements on field cooling have revealed that $|\chi_{\rm 2K}|$ regarded as corresponding to the SC volume fraction in a sample increases with increasing $x$ up to $x=0.089$, exhibits the minimum around $x=0.115$ and increases again up to $x=0.178$. 
The specific-heat measurements have shown that $\gamma$ in the ground state decreases with increasing $x$ and tends to be saturated around the optimally doped regime of $x\sim0.18$. 
These indicate that both the SC volume fraction in a sample, which is one of conclusive properties to confirm a phase separation, and the density of states of quasiparticles at the Fermi level significantly depend on $x$. 
These results strongly suggest that a phase separation into SC and normal-state regions takes place in the underodped regime of LSCO as well as in the overdoped regime. 
Therefore, it is concluded that the phase separation, that is, the electronic inhomogeneity is a bulk property of a sample and is possibly a generic feature of the hole-doped high-$T_{\rm c}$ cuprates. 
The present results imply that $T_{\rm c}$ changes with $p$ depending on the ratio between the SC and non-SC regions, that is, $T_{\rm c}$ increases accompanied by the gradual evolution of the SC coherence between SC regions with increasing $p$. 
Accordingly, the present results imply an importance of the inhomogeneous SC state for understanding of the mechanism of the high-$T_{\rm c}$ superconductivity.

\section*{Acknowledgments}
Fruitful discussions with H. Tsuchiura and Y. Yanase are gratefully acknowledged. 
We are indebted to M. Ishikuro for his help in the ICP analysis. 
The $\chi$ measurements were carried out at the Center for Low Temperature Science, Tohoku University. 
The EPMA analysis was supported by Y. Murakami in the Advanced Research Center of Metallic Glasses, Institute for Materials Research (IMR), Tohoku University and K. Kudo in IMR, Tohoku University.
This work was supported by the Iketani Science and Technology Foundation and also by a Grant-in-Aid for Scientific Research from the Ministry of Education, Culture, Sports, Science and Technology, Japan.

\end{document}